\documentclass[12pt]{article}

\usepackage[latin1]{inputenc}
\usepackage[T1]{fontenc}
\usepackage[english]{babel}
\usepackage{graphicx}
\usepackage{float}

\usepackage{tikz}
\usepackage{[caption}
\usetikzlibrary{arrows}
\usetikzlibrary{decorations.markings}
\usetikzlibrary{decorations.pathmorphing}

\usepackage{times}
\usepackage{graphics}

\usepackage{amsmath}
\usepackage{hyperref}
\usepackage{hhline}
\usepackage{subfig}
\usepackage{color}
\usepackage[all]{hypcap}

\graphicspath{{figures/}}
\newcommand {\ra}        {\rightarrow}
\newcommand {\mumemconv}[1][A] {\mbox{$\mu^- \textrm{#1} \rightarrow e^- \textrm{#1}$}}
\newcommand {\mumepconv}[1][A] {%
  \def\ArgI{{#1}}
  \mumepconvRelay
}
\newcommand \mumepconvRelay[1][A]  {\mbox{$\mu^- \textrm{\ArgI} \rightarrow e^+ \textrm{#1}$}}

\newcommand {\Pb}[1]     {\mbox{$\rm ^{#1}Pb$}}                 
\newcommand {\Au}[1]     {\mbox{$\rm ^{#1}Au$}}                 
\newcommand {\Pt}[1]     {\mbox{$\rm ^{#1}Pt$}}                 
\newcommand {\Ir}[1]     {\mbox{$\rm ^{#1}Ir$}}                 
\newcommand {\kmax}      {\mbox{$k_{\rm max}$}}
\begin{document}

\begin{titlepage}

  
  \begin{center}
    {\Large \bf Search for  $\mu^- \rightarrow e^+$  conversion: what can be learned
      from the SINDRUM-II positron data on a gold target} 
    \vspace{0.5cm}
    
    M. MacKenzie (Northwestern University), P. Murat(FNAL)

    \vspace{0.5cm}
  \end{center}

  \begin{abstract}
    In their 2006 paper setting the current limit on \mumemconv\ conversion search
    on a gold target \cite{sindrum_ii:Bertl2006}, the SINDRUM-II collaboration published,
    along with the electron momentum distribution, the momentum distribution of
    reconstructed positrons. Near the positron spectrum endpoint, there is a statistically
    significant excess of observed events over the expected background.
    We estimate that in the region 88 MeV/c < p < 95 MeV/c there are 13 events with an 
    expected background of about 1-1.5 event, which has not been discussed by the authors.
    Those 13 events form a bump with a width consistent with the experimental 
    resolution, making one think of a $\mu^- \rightarrow e^+$ conversion signal.
    However, the reconstructed position of the bump is about 1 MeV/c,
    or $\sim\ 4\sigma_p$, lower than the expected position of the $\mumepconv[Au][Ir]$ signal,
    which strongly discourages the exotic interpretation.
    
    The excess, however, could be due to an exclusive dipole radiative muon capture (RMC)
    transition $^{197} \rm Au(GS) ~\ra~ ^{197} \rm Pt(GS)$ with the branching fraction
    of about $2\cdot 10^{-4}$. Such a transition would not be resolved by the existing
    RMC measurements. 
    We conclude that the exclusive RMC transitions could significantly modify
    the positron spectrum near the kinematic endpoint, and to fully exploit the physics
    potential of the upcoming experiments such as Mu2e and COMET, a better theoretical
    understanding of the endpoint of the RMC spectrum on nuclei is needed.
    A high-resolution measurement of the RMC photon spectra needs to be carried out
    and compared to the theoretical predictions. Without that,
    the sensitivity of the searches for $\mumepconv$ and $\mumemconv$ might be severely
    limited by unknown probabilities of RMC transitions to the exclusive low-lying
    states of the daughter nuclei.
  \end{abstract}

\end{titlepage}

{\tableofcontents}



\newpage
\section { Introduction}

The most stringent search limits on the processes of \mumepconv\ and \mumemconv\ 
on nuclei come from the SINDRUM-II experiment :

$$
Br(\mumepconv[Ti][Ca]) < 1.7\cdot 10^{-12}(GS) ~~~\text{\cite{sindrum_ii:Kaulard1998}}
$$
and 
$$
Br(\mumemconv[Au]) < 7\cdot 10^{-13}  ~~~\text{\cite{sindrum_ii:Bertl2006}},
$$

The data on the gold target have been collected in 2000, after \cite{sindrum_ii:Kaulard1998} 
had been published, and the total total flux of stopped muons collected by SINDRUM-II on an Au
target is about 1.5 times higher than the corresponding number of muon stops on Ti,
$(4.37\pm0.32)\cdot 10^{13}$  \cite{sindrum_ii:Bertl2006} vs $(2.95\pm0.13)\cdot 10^{13}$ \cite{sindrum_ii:Kaulard1998} muon stops respectively.


For a low background experiment, a larger number of stopped muons usually leads to a better
experimental sensitivity, however, results of the search for \mumepconv\ conversion on an Au
target have not been published. Nevertheless, along with the 
momentum spectrum of electrons, SINDRUM-II also shows the positron data 
\cite{sindrum_ii:Bertl2006}-
see Figure  ~\ref{fig:sindrum_ii_2006_fig_11}.

\vspace{0.2in}
\begin{tikzpicture}
  \hspace*{-1cm}%
  \node[anchor=south west,inner sep=0] at (0,0.) {
    \makebox[\textwidth][c] {
      \includegraphics[width=0.8\textwidth, trim = 0 0 0 30,clip]{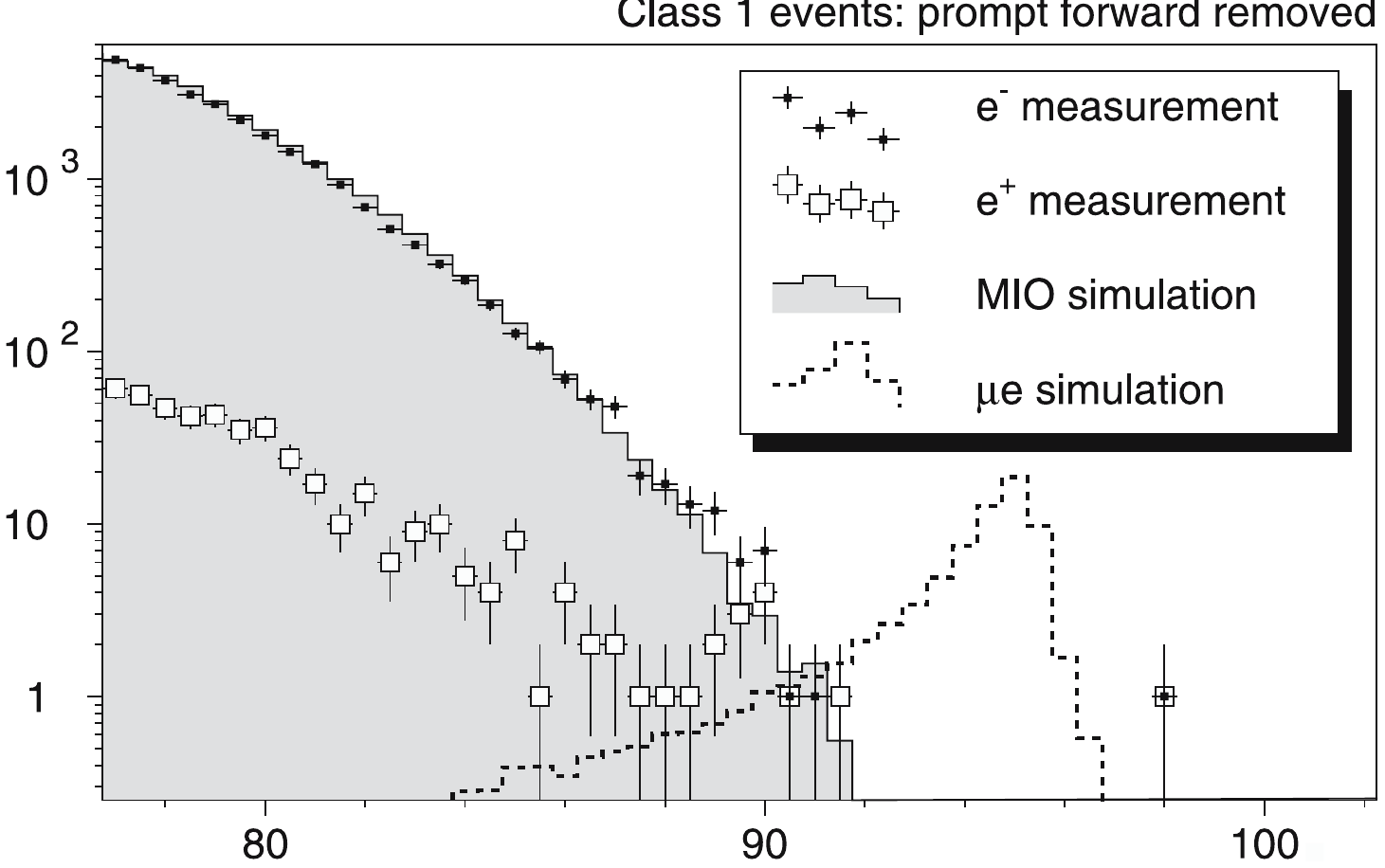}
    }
  };
\end{tikzpicture}
{\captionof{figure} {
  \label{fig:sindrum_ii_2006_fig_11}
  SINDRUM-II $e^-$ and $e^+$ spectra on an \Au{197}\ target from \cite{sindrum_ii:Bertl2006}.
}
}
\vspace{0.1in}

The positron momentum distribution in Figure \ref{fig:sindrum_ii_2006_fig_11}
has a  bump near the spectrum endpoint. 
This note presents an attempt to understand the origin of the bump and its implications
for upcoming searches for \mumepconv\ conversion by the Mu2e and COMET experiments.

\newpage
\section { Approach}

We start by developing a parameterized model of the SINDRUM-II detector response
and tuning the model parameters to describe the electron data from \cite{sindrum_ii:Bertl2006}.
The tuning procedure assumes that the measured electron spectrum is dominated 
by electrons produced in muon decays in orbit (DIO). This assumption doesn't take
into account electrons from radiative muon capture (RMC). Those electrons are definitely
present in the data, and Figure ~\ref{fig:sindrum_ii_2006_fig_11} shows that in the
vicinity of 90 MeV/c their contribution is non-negligible and needs to be taken into account.

However, as reference \cite{sindrum_ii:Bertl2006} doesn't show the momentum spectrum
of RMC electrons, which, due to the Compton scattering, is different from the RMC positron
spectrum, we simply acknowledge the presence of the RMC contribution in the SINDRUM-II
electron data, but don't take any action. 

To model the electron data, SINDRUM-II used the DIO spectrum calculated for the muonic Pb atom,
tabulated in \cite{Watanabe:1993}. The spectrum was corrected for the difference in the
muon binding energies in muonic atoms of \Au{197} and \Pb{208}. The parameterized model
of the SINDRUM-II detector response developed in Section \ref{sec:detector_response} uses
this DIO spectrum as input. 

We next assume that the detector response - resolution and efficiency - is the same for electrons
and positrons of the same momentum, and use the model of the SINDRUM-II detector response,
tuned on electrons, to describe the positron spectrum from \cite{sindrum_ii:Bertl2006}.
The spectrum is dominated by positrons coming from conversions of RMC photons in the
detector material as well as from the direct production of $e^+e^-$ pairs in the process
of nuclear muon capture.

To predict the momentum spectrum of positrons, one needs to know the energy distribution
of photons produced in radiative muon capture on gold.   
We assume that for energies at least a few MeV below the spectrum endpoint, the RMC photon
spectrum can be described by the closure approximation model \cite{RoodTolhoek:1965}
and we fit the positron spectrum below 88 MeV/c to determine the model parameter $\kmax$ 
defining the endpoint and shape of the photon energy spectrum. After that, we predict
the expected RMC contribution in the positron spectrum above 88 MeV/c and quantify the
observed excess of events in that region.

Finally, we discuss if the observed excess is consistent with the signal expected from
the exotic $\mumepconv[Au][Ir]$ transition.


\newpage
\section {Parameterization of the SINDRUM-II detector response}
\label{sec:detector_response}

\subsection{Momentum resolution}

Figure ~\ref{fig:dio_1} overlays the electron momentum spectrum measured by SINDRUM-II
and the DIO spectrum on \Pb{208} calculated in \cite{Watanabe:1993} and shifted by 0.5 MeV/c
to account for the difference in the muon binding energies in \Pb{208} and \Au{197}. 
Both spectra are rapidly falling with momentum and have very different slopes.
The sign of the difference is easy to understand: in the case of a steeply falling spectrum,
the finite experimental resolution smears the spectrum and makes it less steep.

\vspace{0.2in}
\begin{tikzpicture}
  \node[anchor=south west,inner sep=0] at (0,0.) {
    \makebox[\textwidth][c] {
      \includegraphics[width=1.0\textwidth]{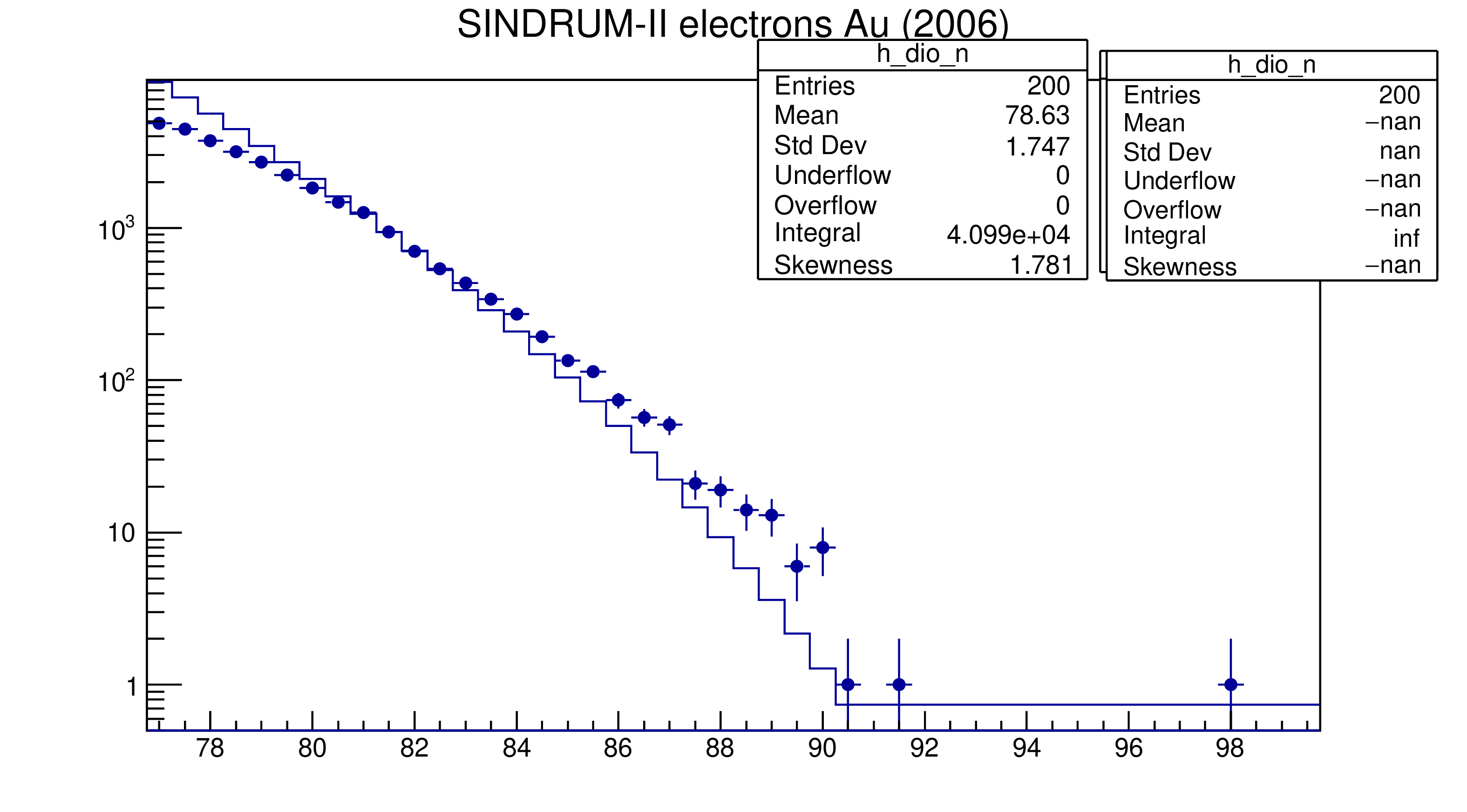}
    }
  };
\end{tikzpicture}
{
  \captionof{figure} {
    \label{fig:dio_1}
    SINDRUM-II electron spectrum (points with the error bars) overlaid with the DIO spectrum
    from \cite{Watanabe:1993}.
    The theoretical spectrum, shown as a histogram, is normalized to the same area as the data
    for p > 80 MeV/c.
  }
}
\vspace{0.2in}

We make a simplifying assumption that the detector momentum response function is symmetric and
can be described by a single Gaussian with a mean of 0. To choose the optimal value of the 
resolution parameter, $\sigma_P$, we vary its value in the range [1, 3.5] MeV/c, convolve
the theoretical DIO spectrum with the resolution function, and use the resulting distribution
to fit the SINDRUM-II electron spectrum. The fit has one parameter - the normalization. 
The fit $\chi^2$ dependence on $\sigma_P$ is shown in Figure ~\ref{fig:ana_step1_fit_sigma.png}.
The best value of the resolution parameter, $\sigma_P = 2.0 \pm 0.1$ MeV/c, is significantly
higher than the SINDRUM-II momentum resolution, which is about 0.5 MeV/c.
The difference is most likely due to the contribution of RMC electrons, unaccounted for
in the fitting procedure. As one can see from Figure ~\ref{fig:sindrum_ii_2006_fig_11},
for momenta close to 90 MeV/c, the RMC contribution can not be ignored. As the RMC spectrum
falls less steeply than the DIO spectrum, ignoring the RMC contribution should result in a 
larger value of $\sigma_P$ returned by the fit. 

\vspace{0.2in}
\begin{tikzpicture}
  \node[anchor=south west,inner sep=0] at (0,0.) {
    \makebox[\textwidth][c] {
      \includegraphics[width=0.9\textwidth, trim = 0 0 0 200, clip]{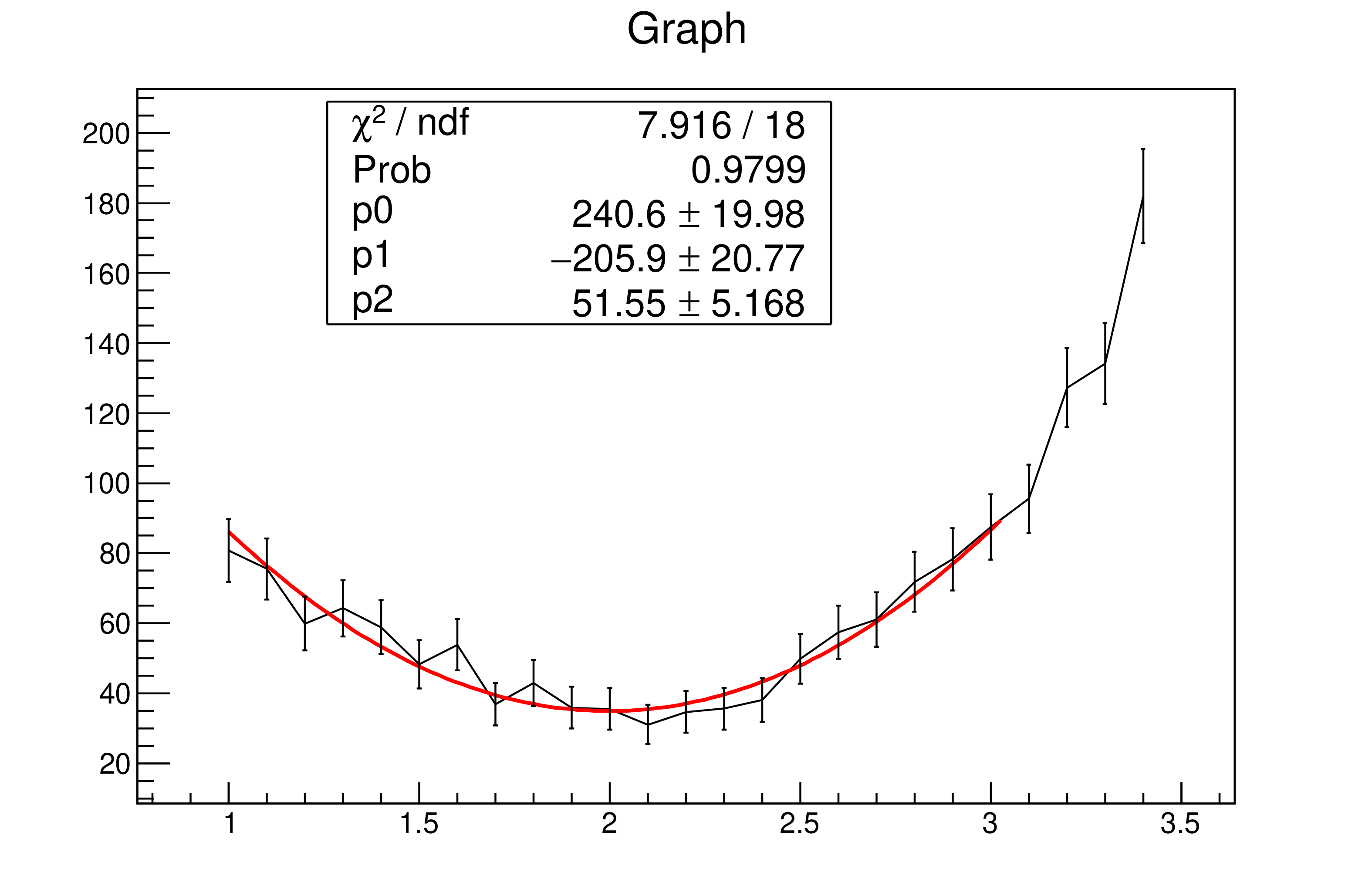}
    }
  };
\end{tikzpicture}
{\captionof{figure} {
  \label{fig:ana_step1_fit_sigma}
  $\chi^2$ of the fit of the SINDRUM-II DIO electron spectrum with the theoretical distribution
  convolved with a Gaussian with resolution $\sigma_P$ as a function of $\sigma_P$.
}}
\vspace{0.2in}

\subsection{Tracking efficiency}

We assume that the SINDRUM-II tracking efficiency is flat for momenta p > 80 MeV/c.
Normalizing the DIO spectrum convolved with a Gaussian resolution function with
$\sigma_P = 2.0$ MeV/c to the electron data in the region p > 80 MeV/c,
and dividing the data spectrum by the resulting distribution gives the ``efficiency''
dependence on the track momentum. The resulting dependence is shown
in Figure ~\ref{fig:ana_step1_efficiency}, it is well described by a function,
constant above 80 MeV/c and falling linearly below 80 MeV/c.

\begin{tikzpicture}
  \node[anchor=south west,inner sep=0] at (0,0.) {
    \makebox[\textwidth][c] {
      \includegraphics[width=0.9\textwidth]{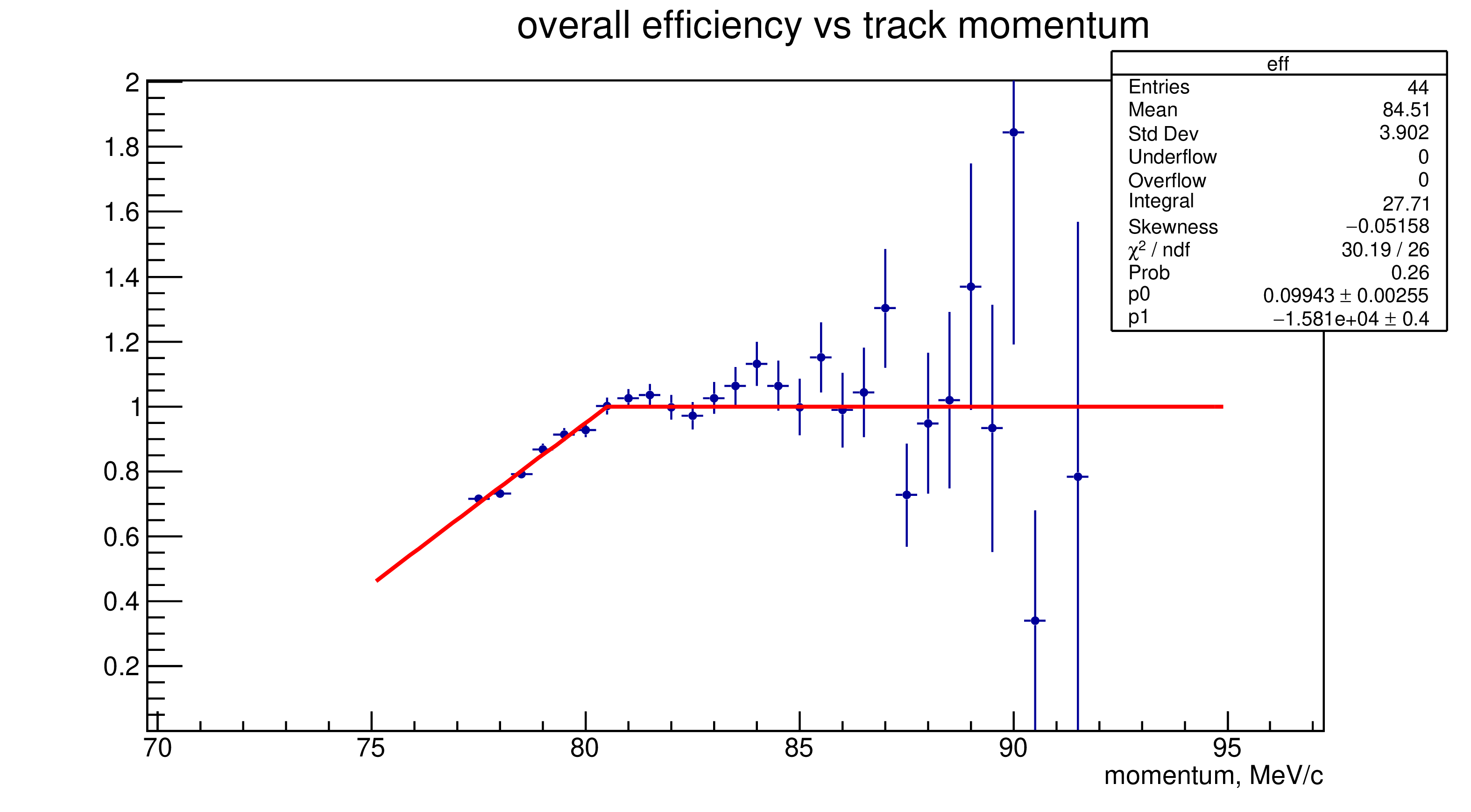}
    }
  };
\end{tikzpicture}
{\captionof{figure} {
  \label{fig:ana_step1_efficiency}
  Parameterization of the SINDRUM-II efficiency vs the track momentum.
  Definition of efficiency includes all components - trigger, reconstruction, and selection.
  Overall normalization is chosen such that efficiency is equal to one for p > 80 MeV/c.
}}
\vspace{0.2in}

This step concludes tuning of the detector response. Figure \ref{fig:ana_step1_best_dio_fit}
shows the description of the SINDRUM-II electron data of \cite{sindrum_ii:Bertl2006}
with the tuned response - the quality of description is surprisingly good,
better than one might expect from such a simplistic model.

\vspace{0.2in}
\begin{tikzpicture}
  \node[anchor=south west,inner sep=0] at (0,0.) {
    \makebox[\textwidth][c] {
      \includegraphics[width=0.99\textwidth]{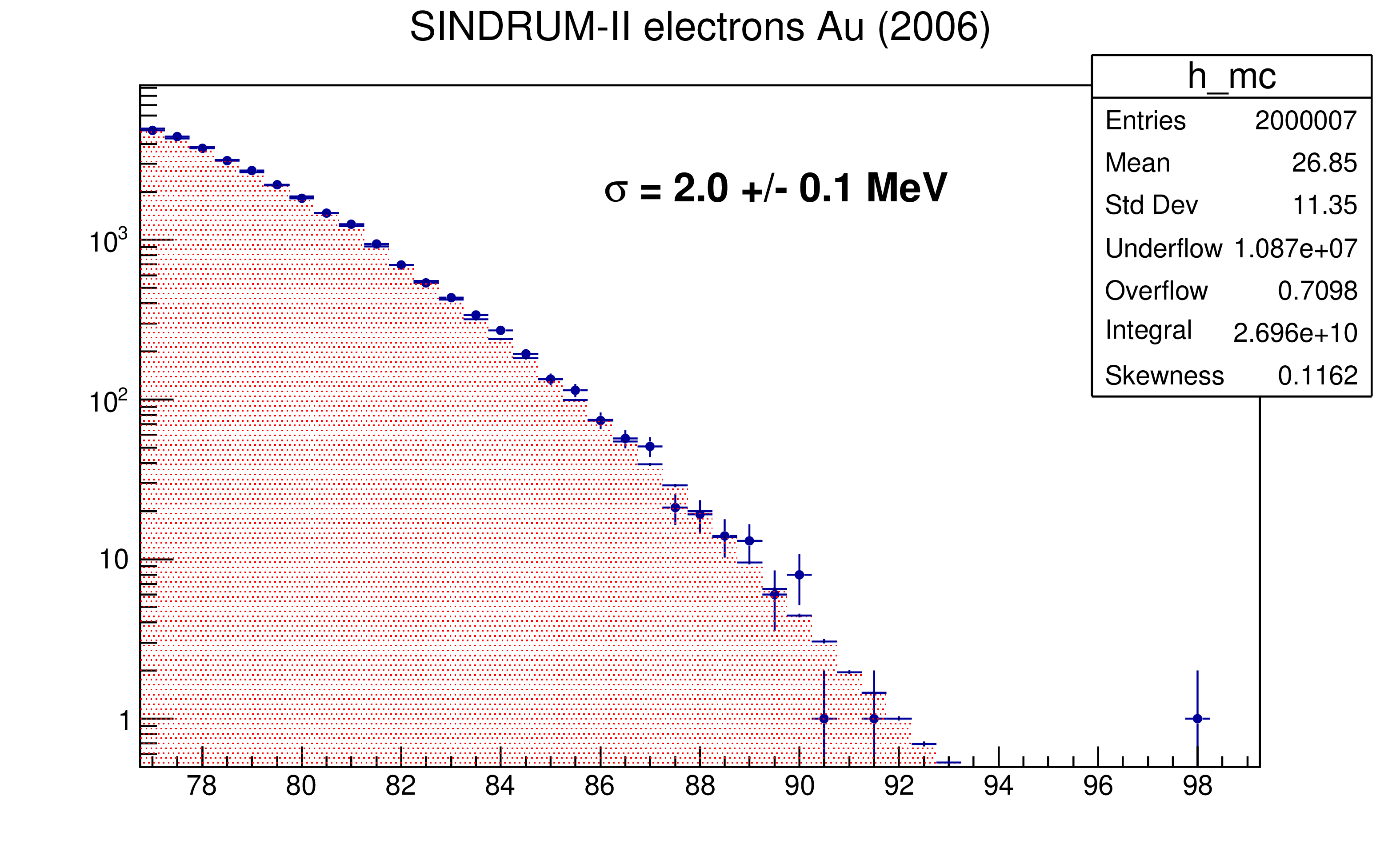}
    }
  };
\end{tikzpicture}
{\captionof{figure} {
  \label{fig:ana_step1_best_dio_fit}
  Description of the electron spectrum on the Au target from \cite{sindrum_ii:Bertl2006}
  with the tuned model of the SINDRUM-II detector response.
}}

\newpage
\section {RMC spectrum and $\kmax$ determination}
\label{sec:kmax_determination}

Measured RMC photon energy spectra on nuclei can be successfully described
within a closure approximation model which predicts the RMC photon
spectra depending on just one parameter - the photon spectrum endpoint:
$$
    \frac{dN}{dE} ~=~ \frac{e^2}{\pi} \frac{\kmax^2}{ m_{\mu}^2} (1 - \alpha) (1-x+2x^2)x(1-x)^2
$$
where E is the photon energy, $\kmax$ is the energy spectrum endpoint, $\rm \alpha = (N-Z)/A$,
and $x = E/\kmax$ \cite{RoodTolhoek:1965}.
Values of $\kmax$ for different nuclei are not predicted, but determined from fits
to the experimental data. Typically, fits return $\kmax$ values significantly, 5-15 MeV/c,
lower than the kinematically allowed limits. 

Internal conversions, or off-shell RMC photons, also contribute to the measured
spectrum. Assuming that the energy distribution of off-shell photons is the same
as the energy distribution of the on-shell photons, and the energy sharing between
the positron and electron is similar in both cases, the contribution of the off-shell
photons should be accounted for by the overall spectrum normalization.

We determine the value of the $\kmax$ parameter by fitting the SINDRUM-II positron spectrum
with a closure approximation spectrum convolved with the resolution ($\sigma_P = 2$ MeV/c)
and the efficiency determined from the fit to the electron spectrum 
(see Figure ~\ref{fig:ana_step1_efficiency}) for a range of $\kmax$ values.
Figure ~\ref{fig:ana_step2_fit_kmax} shows the dependence of the fit $\chi^2$ on $\kmax$,
the best fit corresponds to $\kmax = 88.0 \pm 0.6$ MeV.
With the positron energy losses taken into account (see Section \ref{sec:momentum_scale}),
the value becomes $\kmax = 88.6 \pm 0.6$ MeV. Consistent with the available experimental data,
the maximal photon energy allowed kinematically in a 
$\rm \mu^- + \Au{197}(GS) \rightarrow \nu +\gamma + ^{197}Pt(GS) $ transition
is $E_{max} = 94.3$ MeV, about 5 MeV higher than the \kmax\ value
corresponding to the best fit.
\vspace{0.1in}
\begin{tikzpicture}
  \node[anchor=south west,inner sep=0] at (0,0.) {
    \makebox[\textwidth][c] {
      \includegraphics[width=1.0\textwidth, trim = 180 0 50 125,clip]{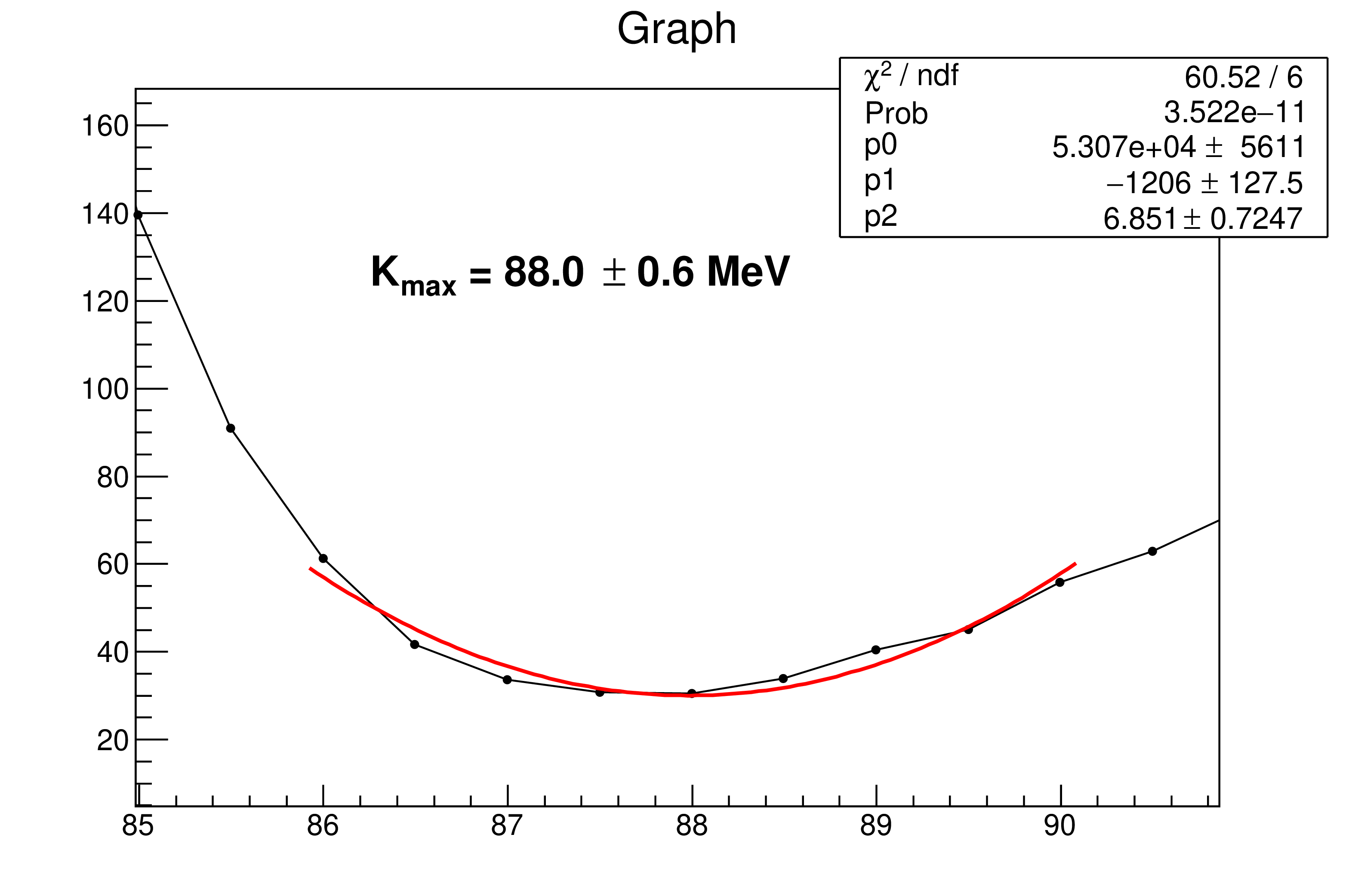}
    }
  };
\end{tikzpicture}
{\captionof{figure} {
  \label{fig:ana_step2_fit_kmax}
  $\chi^2$ of the SINDRUM-II positron spectrum fit with the posiron momentum distribution
  derived from the RMC closure approximation spectrum convolved with the detector response
  as a function of $\kmax$. 
  Used in the fit are events with p < 88 MeV/c.
}}
\vspace{0.1in}


\newpage
\section {SINDRUM-II momentum scale }
\label{sec:momentum_scale}

\begin{figure}[H]
\begin{tikzpicture}
  \node[anchor=south west,inner sep=0] at (0,0.) {
    \makebox[\textwidth][c] {
      \includegraphics[width=0.99\textwidth]{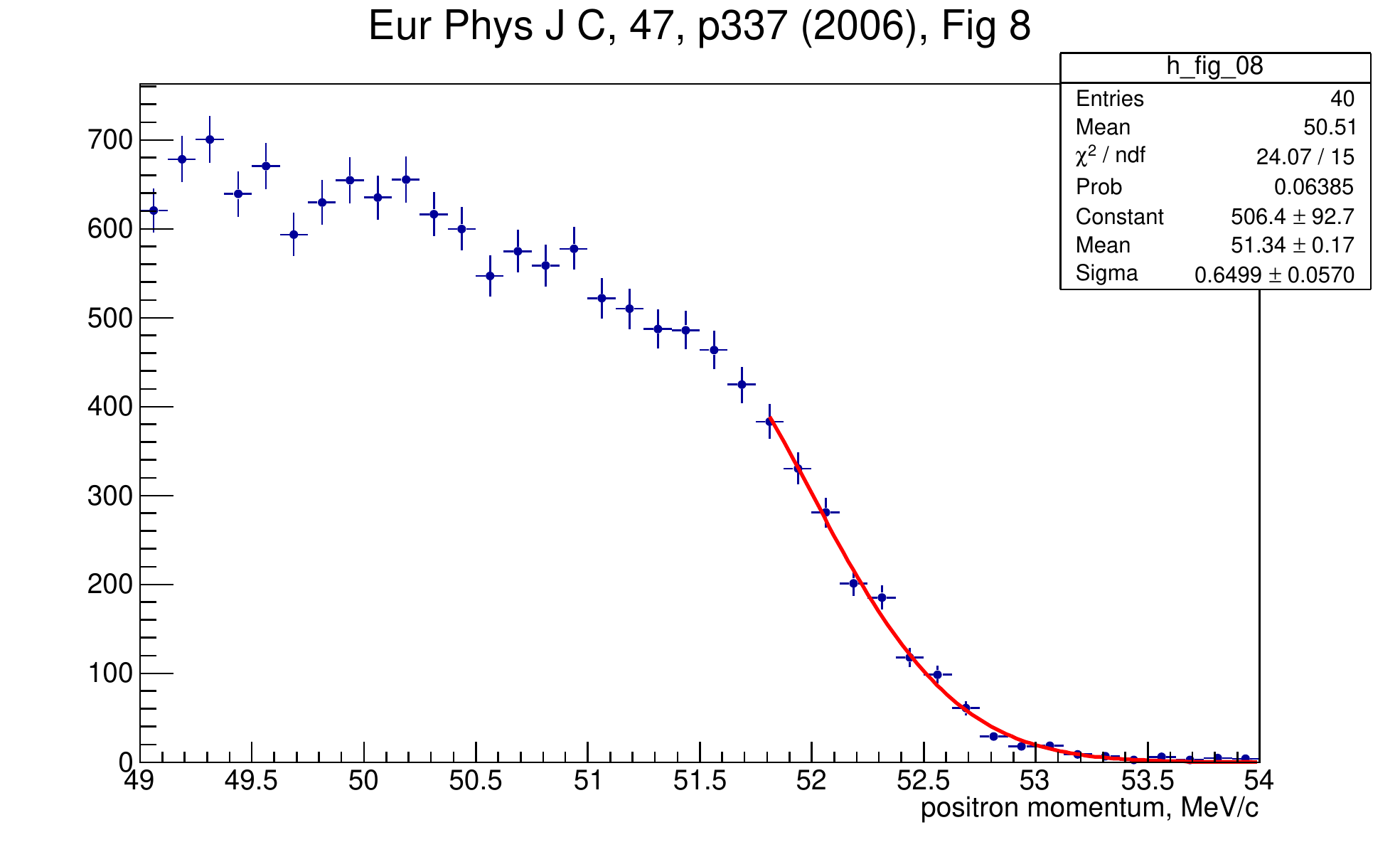}
    }
  };
\end{tikzpicture}
\caption{
  \label{fig:sindrum_ii_fig_08_fit}
  Reconstructed momentum spectrum of positrons from $\mu^+ \rightarrow e^+ \nu \bar{\nu}$
  decays used in \cite{sindrum_ii:Bertl2006} for detector momentum calibration.
}
\end{figure}

In \cite{sindrum_ii:Bertl2006}, the experimental momentum scale has been calibrated
using the edge of the Michel spectrum from muon decays $\mu^+ \rightarrow e^+ \nu \bar{\nu}$
at rest. The calibration was performed with the magnetic field reduced to about 50\%
of the nominal value. The reconstructed positron momentum distribution is shown
in Figure \ref{fig:sindrum_ii_fig_08_fit}.

Although radiative corrections modify the positron spectrum, their impact on the edge
of the Michel spectrum is fairly small, as shown in Figure \ref{fig:mu2e_5645_fig_001_mue3_decay}.
Therefore, the reconstructed position and shape of the edge depend primarily on the energy losses
in the detector and the experimental momentum resolution.

\begin{tikzpicture}
  \node[anchor=south west,inner sep=0] at (0,0.) {
    \makebox[\textwidth][c] {
      \includegraphics[width=0.99\textwidth]{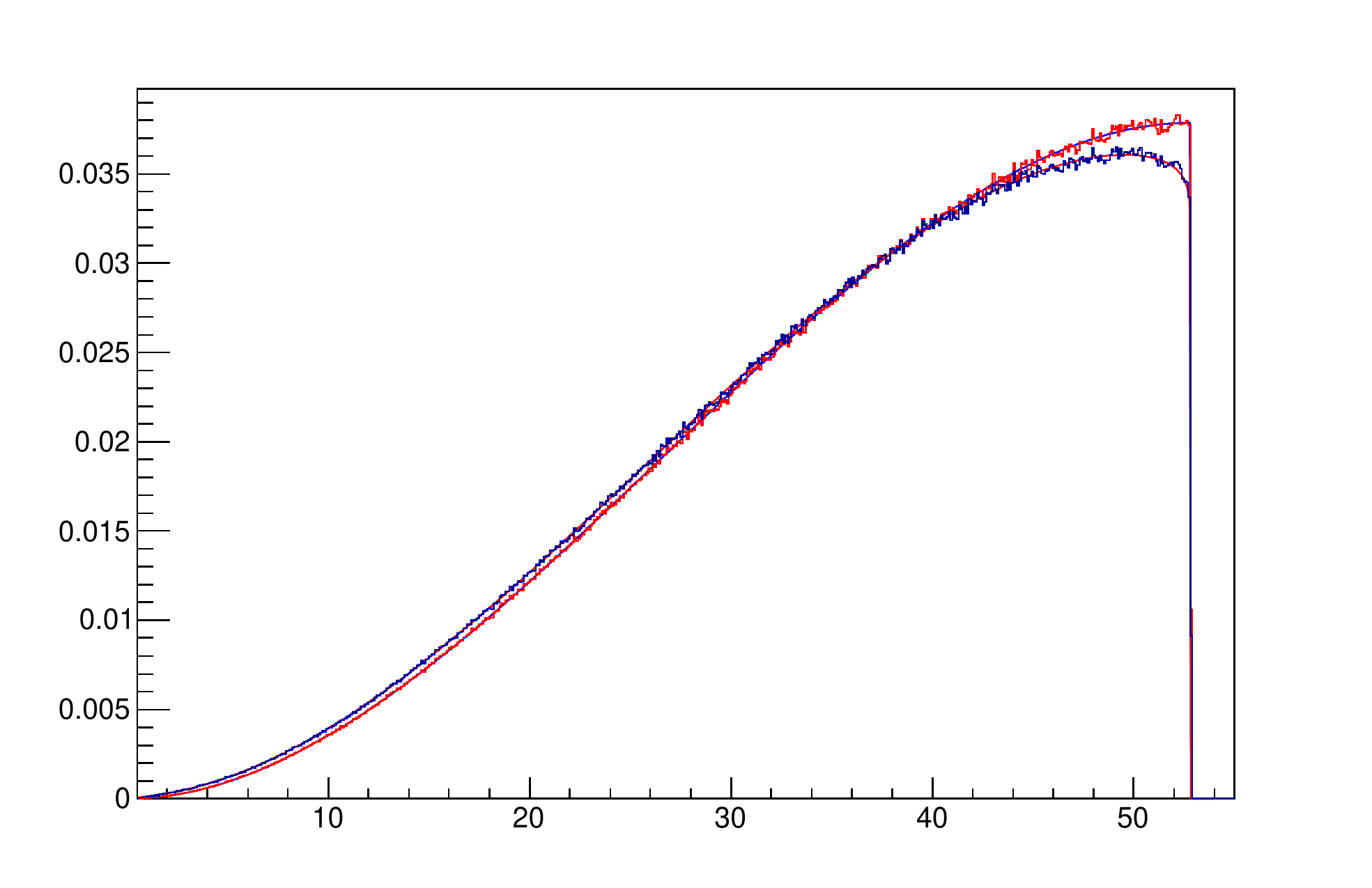}
    }
  };
\end{tikzpicture}
{\captionof{figure} {
  \label{fig:mu2e_5645_fig_001_mue3_decay}
  Theoretical Momentum spectrum of positrons from the $\mu^+ \rightarrow e^+ \nu \bar{\nu}$
  decay. Red: leading order, blue: with radiative corrections taken into account.
}}
\vspace{0.1in}

We expect the positron momentum referred to in the paper
to be the momentum in the first reconstructed point on the trajectory. As such, the
reconstructed Michel edge should be affected by the energy losses in front of the
tracker, as well as the tracker momentum resolution.
According to \cite{sindrum_ii:Bertl2006}, the energy losses in front
of the tracker are due to losses in the Au target (75 mg/cm$^2$) and the wall
of the vacuum chamber (324 mg/cm$^2$), most of which is aluminum and carbon fiber.

\begin{figure} 
\begin{tikzpicture}
  \node[anchor=south west,inner sep=0] at (0,0.) {
    \makebox[\textwidth][c] {
      \includegraphics[width=0.99\textwidth]{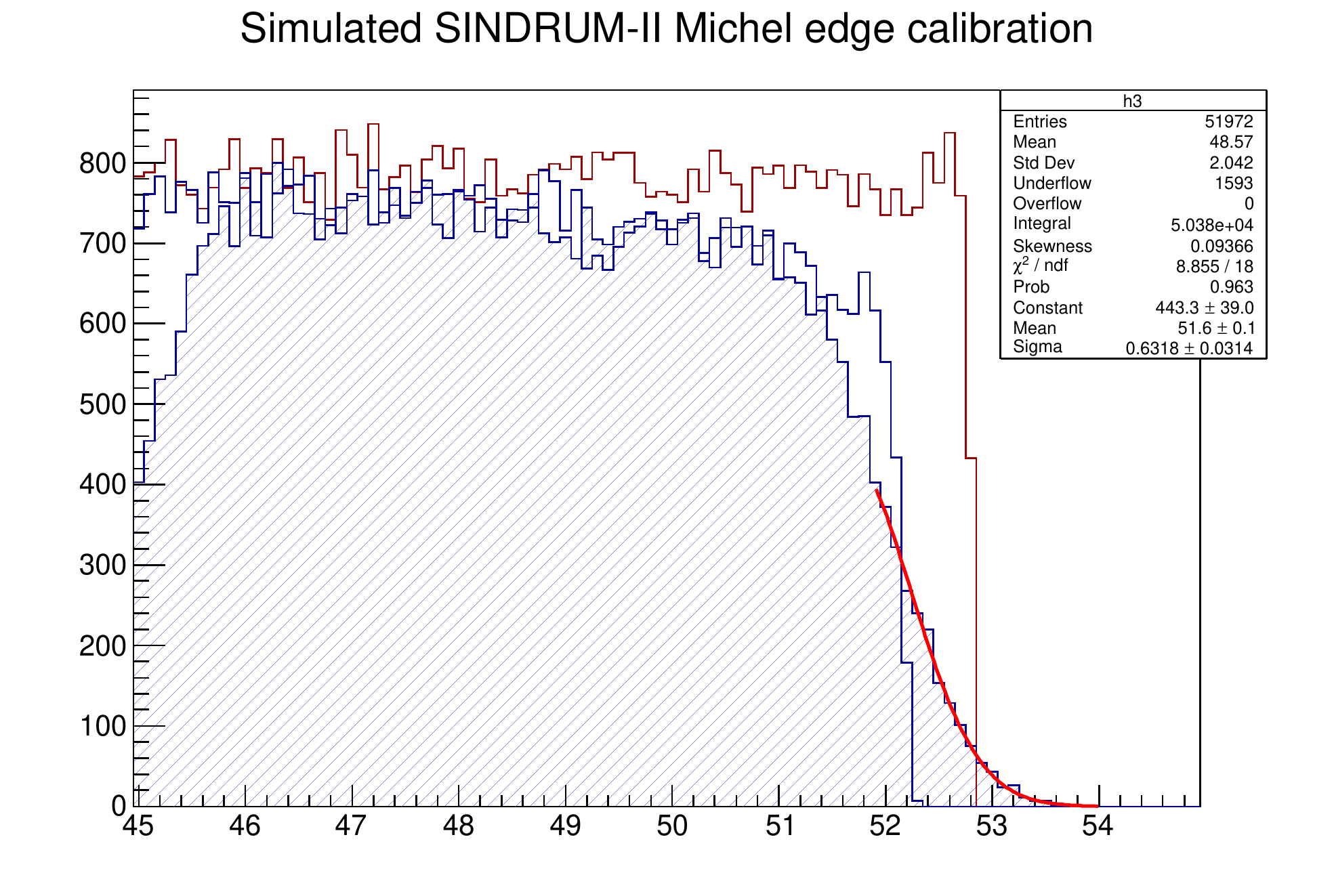}
    }
  };
\end{tikzpicture}
\caption{
  \label{fig:sindrum_ii_michel_calibration}
  A flat spectrum with the right edge at 52.8 MeV is shown in red,
  the same spectrum convolved with the expected SINDRUM-II energy losses is in blue,
  and then this distribution convolved with a Gaussian with $\sigma$ = 0.55 MeV/c is shaded.
}
\end{figure}

To validate our understanding of the SINDRUM-II momentum calibration,
we simulate energy losses of positrons with initial momenta distributed uniformly
in the range [45,52.8] MeV/c in a structure consisting of the two layers described
above. The initial momentum distribution is shown in Figure ~\ref{fig:sindrum_ii_michel_calibration}
in red, and the positron momentum distribution on exit from the vacuum chamber wall is 
shown in blue.
Comparing the two distributions shows that for a 50 MeV positron, the most probable
energy loss in front of the SINDRUM-II tracking chamber is about 0.6 MeV

Shaded in Figure ~\ref{fig:sindrum_ii_michel_calibration} is the positron
momentum distribution which includes the energy losses in the target and the
vacuum chamber wall, and is convolved with a Gaussian with $\sigma = 0.55$ MeV/c.
The value of $\sigma = 0.55$ MeV/c corresponds to the SINDRUM-II momentum resolution
of 1.3 MeV/c FWHM \cite{sindrum_ii:Kaulard1997_Thesis}. The shaded distribution
in Figure ~\ref{fig:sindrum_ii_michel_calibration} double counts the fluctuations
of energy losses, however the impact of double counting is small, and the momentum
edge smearing is dominated by the momentum resolution of the tracker.

The fit of the high-momentum part of the smeared edge with a Gaussian returns
$\sigma = 0.63 \pm 0.03$ MeV/c, in good agreement with $\sigma = 0.65 \pm 0.06$ MeV/c
returned by the fit of the SINDRUM-II spectrum in Figure ~\ref{fig:sindrum_ii_fig_08_fit}.

\newpage
\section {Events above 88 MeV and the \mumepconv\ signal}
\subsection {Positron momentum spectrum}

Figure \ref{fig:ana_step2_ppos_best_fit} overlays the SINDRUM-II positron spectrum
on the \Au{197}\ target with the closure approximation spectrum convolved with the 
parameterized model of the SINDRUM-II detector response.

There are 14 events above 88 MeV/c in the data. Looking at the data above 94 MeV/c,
we can estimate the background due to radiative pion capture (RPC) and cosmics to be
of the order of one event. The closure approximation-based RMC model predicts
about 0.4 events above 88 MeV/c. 

One therefore needs to ask whether the closure approximation reliably describes
the RMC positron spectrum near the endpoint. In particular, the closure approximation
doesn't take into account RMC transitions to the exclusive low-lying states of the 
daughter nucleus.

In the case of the \Au{197} target, a dipole transition \Au{197} $\ra$ \Pt{197}
to the ground state of the Pt nucleus is allowed. Given that the energy splitting
between an electron and a positron from a photon conversion is almost uniform,
the positron spectrum corresponding to such a transition, in a first order approximation,
should be flat. Taking into account the energy losses, the positron momentum spectrum
could extend up to 93.2 MeV/c.
Therefore, if it existed, such a transition could reduce the tension between the background
model and the data. Below, we estimate the probability of such a transition corresponding
to 13 RMC events in the region 88 MeV/c < $p_{e^+}$ < 93.2 MeV/c.

\vspace{0.1in}
\begin{tikzpicture}
  \node[anchor=south west,inner sep=0] at (0,0.) {
    \makebox[\textwidth][c] {
      \includegraphics[width=1.0\textwidth]{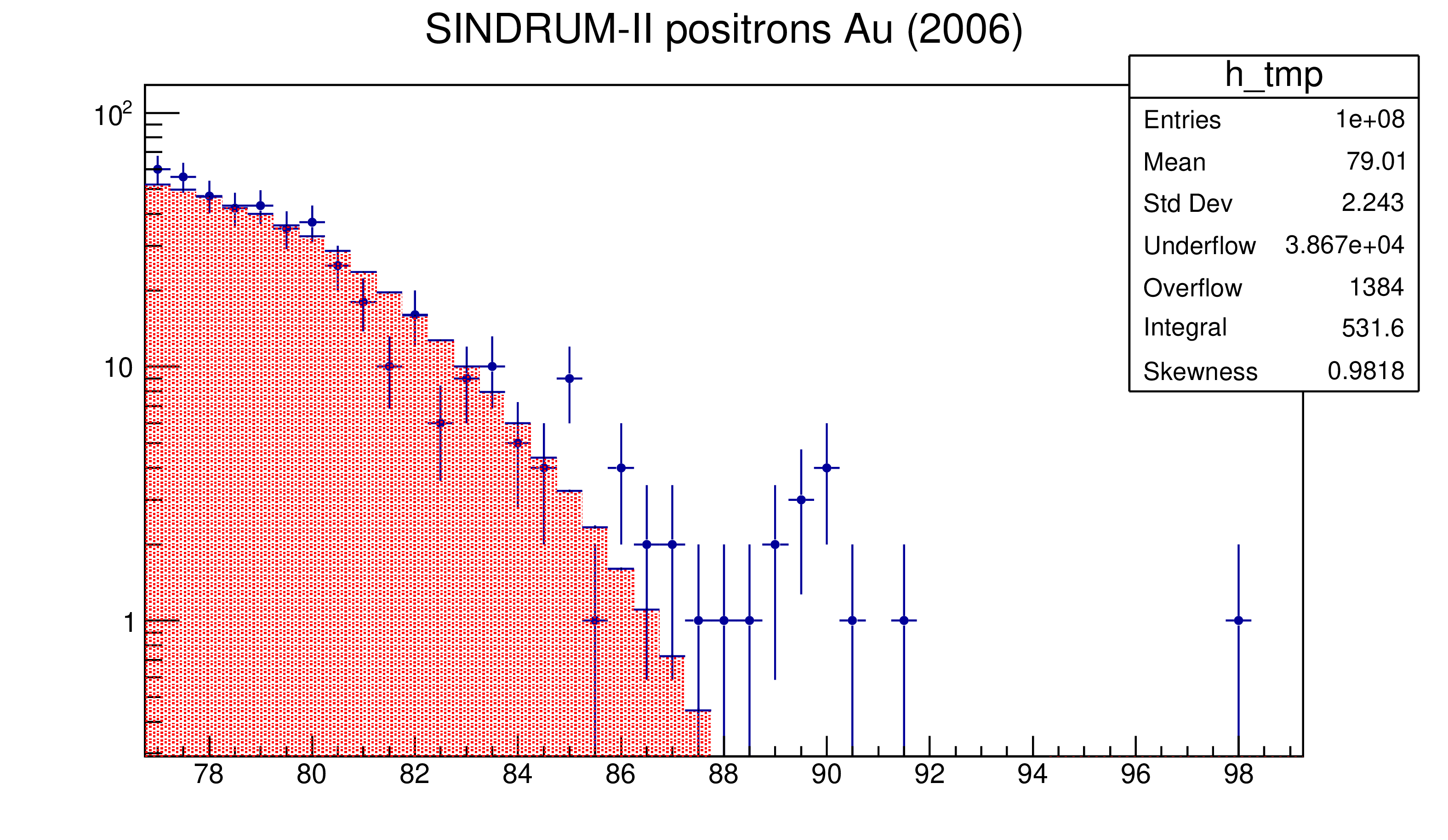}
    }
  };
\end{tikzpicture}
{\captionof{figure} {
  \label{fig:ana_step2_ppos_best_fit}
  Fit of the SINDRUM-II positron spectrum with the closure approximation
  spectrum convolved with the parameterized model of the SINDRUM-II detector
  response. The closure approximation parameter $k_{max} = 88$ MeV/c.
}}
\vspace{0.1in}

\subsection{Final states with a broken down daughter nucleus}

\begin{figure}
  \begin{tikzpicture}
    \node[anchor=south west,inner sep=0] at (-2.0,0.0) {
      \includegraphics[width=0.6\textwidth]{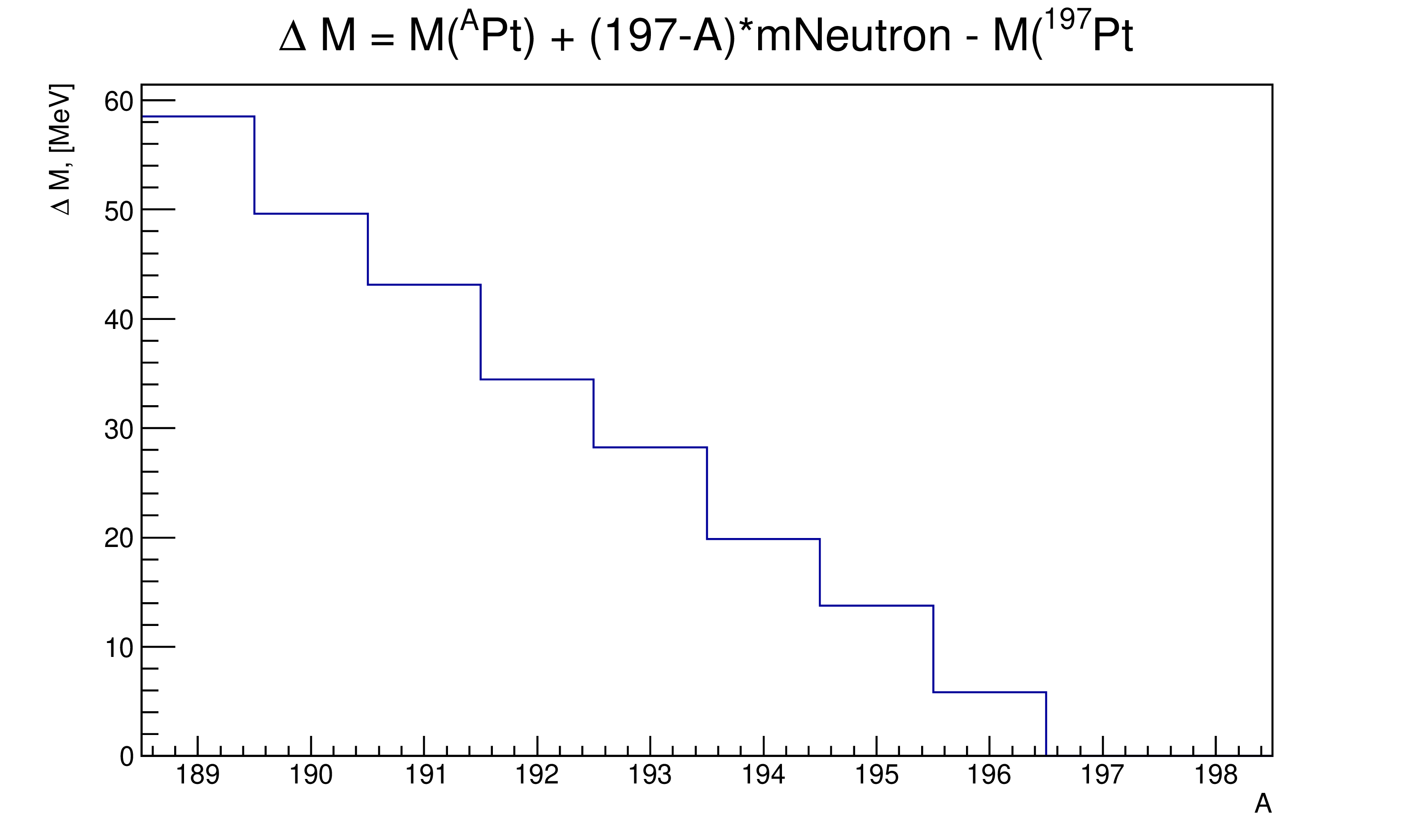}
    };
    \node[anchor=south west,inner sep=0] at (6.0,0.0) {
      \includegraphics[width=0.6\textwidth]{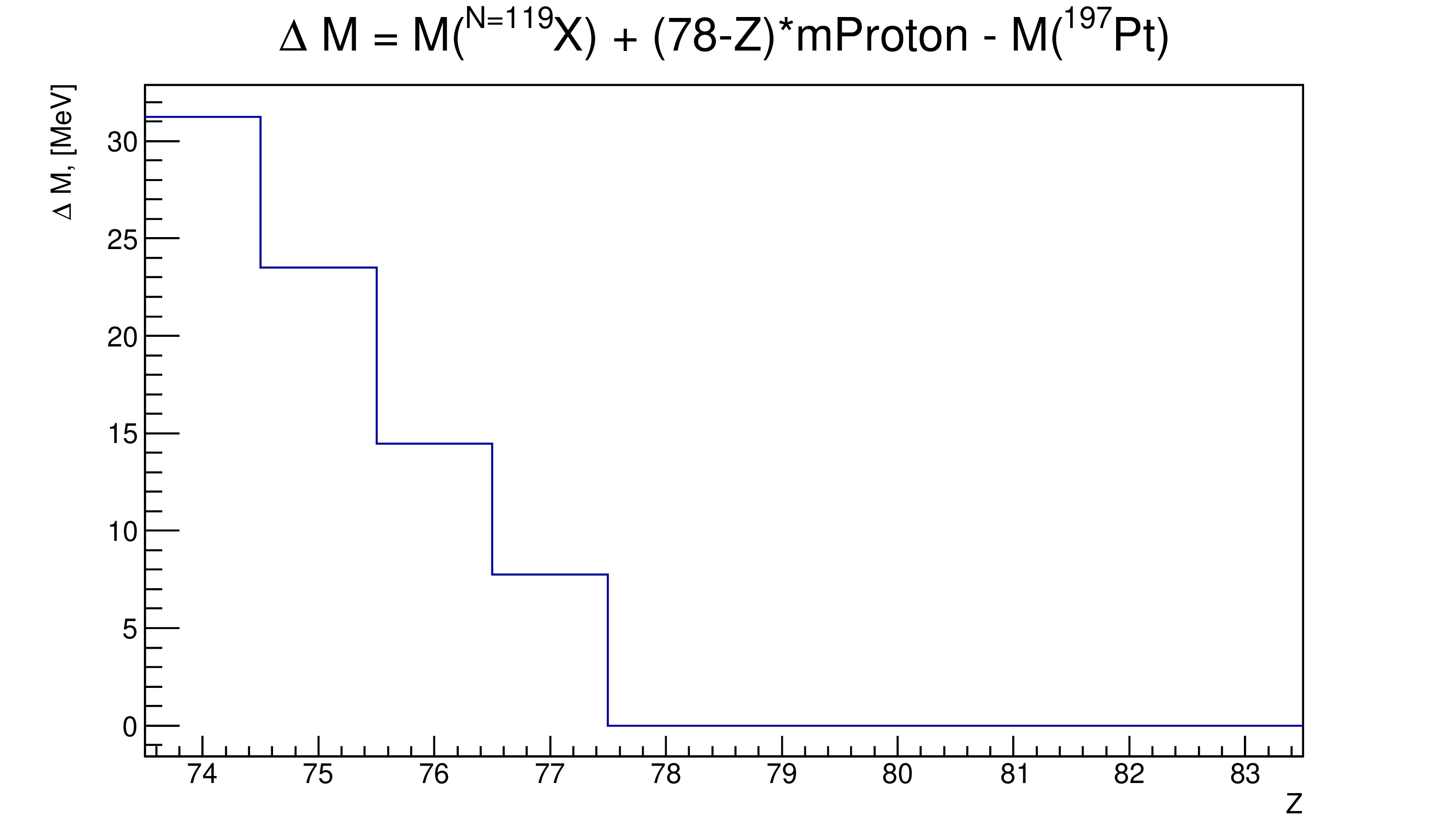}
    };
  \end{tikzpicture}
  \caption{
    \label{fig:pt197_dmass}
    Mass differences between the final state with the broken down daughter nucleus
    and $^{197} \rm Pt$ for different breakdown scenarios of \Pt{197} nucleus.
    Isotope masses taken from \cite{NuclearData:2017}.
    The final state with the \Pt{197}\ nucleus always has the lowest
    mass and, therefore, the highest photon spectrum endpoint.
  }
\end{figure}

As a daughter nucleus produced in the process of radiative muon capture can
break down and emit one or several protons or neutrons, one could also ask
whether photons from transitions of
$$
\mu + \Au{197} \rightarrow \gamma + \nu + ^{197-k}Pt + k \rm ~neutrons
$$
or 
$$
\mu + \Au{197} \rightarrow \gamma + \nu + ^{197-k}A_{Z=78-k} + k \rm ~protons
$$

\noindent could produce positrons with a higher energy spectrum endpoint than the 
$\mu +$ $\Au{197}$ $\ra$ $\gamma + \nu +$ $\Pt{197}$ transition.
Figure ~\ref{fig:pt197_dmass} shows the distributions of mass differences
between the final states consisting of ($^{197-k} \rm Pt\ + k$ neutrons) and 
($^{197-k}A_{Z=78-k} + k$ protons),
and the ground state of the $^{197} \rm Pt$ nucleus.
Based on those distributions one can conclude that events with the most energetic photons
and, as such, most energetic positrons, should correspond to transitions with the 
$^{197} \rm Pt$
nucleus in the final state, with a reconstructed positron momentum endpoint of 93.2 MeV/c.
As the positron momentum spectrum extends up to about 92 MeV/c, final states with massess higher
than the mass of the $^{197} \rm Pt$ ground state by $\sim 1.2$ MeV/c$^2$ couldn't contribute
to the excess.

\subsection {Expected $\mumepconv$ signal in the SINDRUM-II detector}

One could ask whether the excess of events on the high-momentum tail of the positron
momentum distribution is consistent with the signal expected from $\mu^- \ra e^+$
conversion on gold. Figure \ref{fig:sindrum_ii_ce_signal} shows the expected $\mu^- \ra e^-$
signal in the SINDRUM-II detector, digitized from Figure 11 of \cite{sindrum_ii:Bertl2006}. 
The conversion peak has its maximum at 95.0 MeV/c, consistent with the most probable energy
losses of 0.6 MeV in front of the tracker, discussed in Section \ref{sec:momentum_scale}.
The expected position of the $\mumepconv[Au][Ir]$ conversion signal is about 3.9 MeV/c 
lower than for the $\mumemconv[Au]$ signal; the difference is small enough to not affect 
the momentum resolution.
\vspace{0.1in}
\begin{tikzpicture}
  \node[anchor=south west,inner sep=0] at (0,0.) {
    \makebox[\textwidth][c] {
      \includegraphics[width=0.99\textwidth]{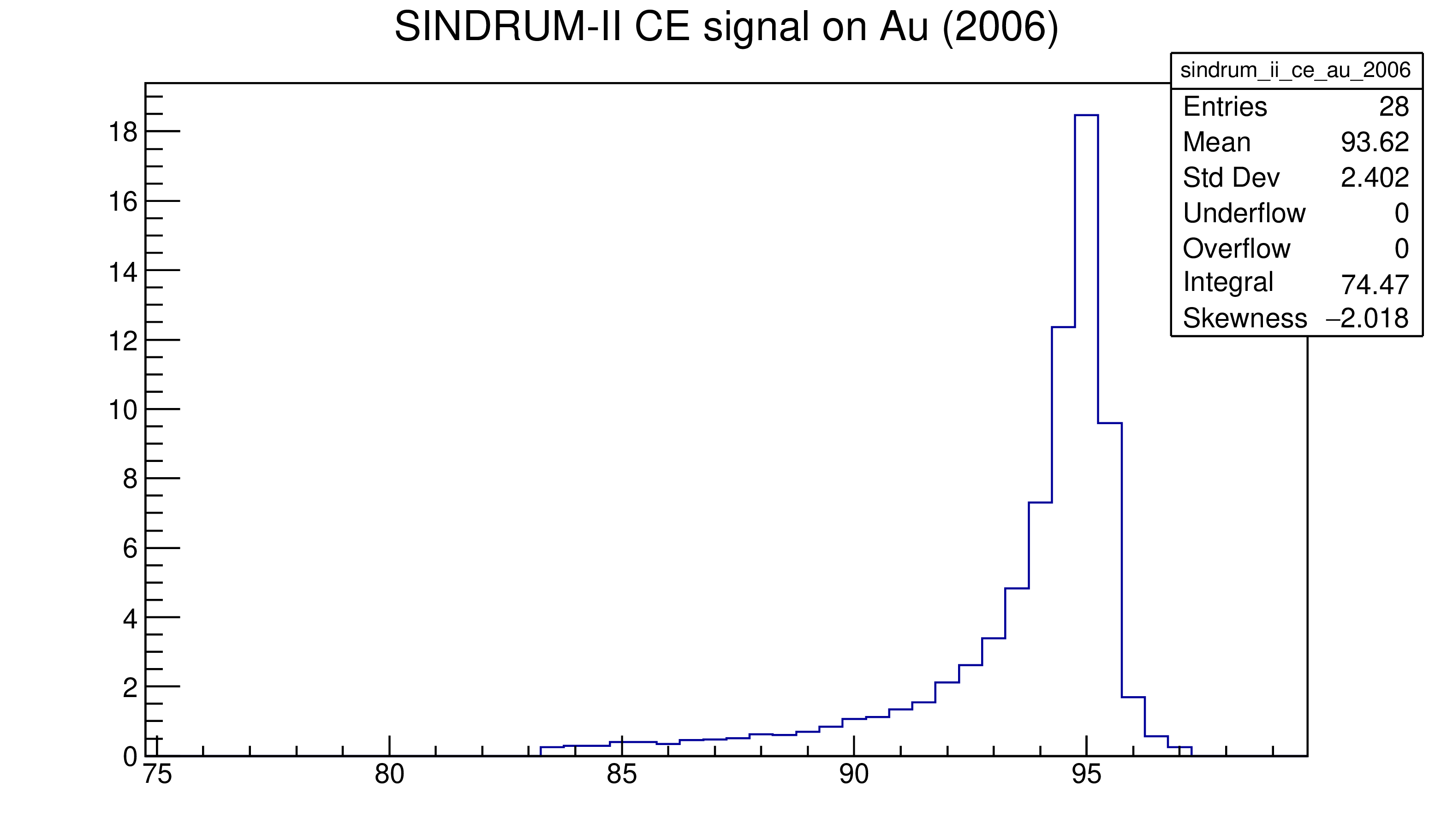}
    }
  };
\end{tikzpicture}
{\captionof{figure}{
  \label{fig:sindrum_ii_ce_signal}
  The expected position and shape of the $\mu^- \rightarrow ~e^-$ conversion signal on Au
  in the SINDRUM-II detector.
}}
\vspace{0.1in}

\begin{figure}
\begin{tikzpicture}
  \node[anchor=south west,inner sep=0] at (0,0.) {
    \makebox[\textwidth][c] {
      \includegraphics[width=1.0\textwidth]{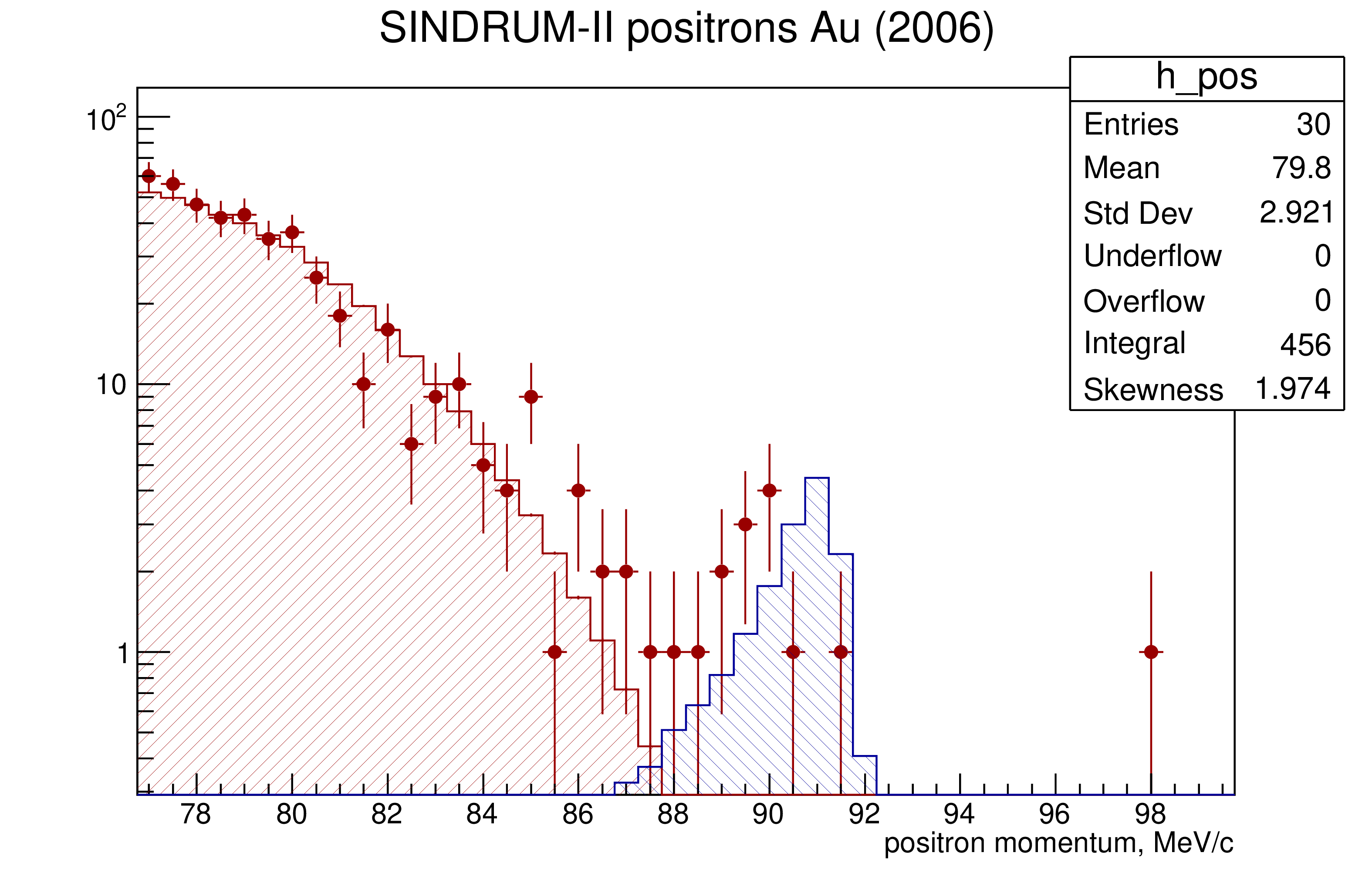}
    }
  };
\end{tikzpicture}
\caption {
  \label{fig:ana_step2_sindrum_positron_best_fit_signal}
  SINDRUM-II positron spectrum overlaid with the expected background from RMC
  and a signal from \mumepconv[Au][Ir]\ normalized to 20 events. Normalization of the
  signal is chosen to guide the eye.
}
\end{figure}

We therefore assume that the expected position and the shape of the positron signal from
$\mu^-\rightarrow e^+$ conversion reconstructed in the SINDRUM-II detector can be reproduced
by moving the $\mu^- \rightarrow e^-$ signal shown in Figure ~\ref{fig:sindrum_ii_ce_signal}
down by 4 MeV/c.
The expected \mumepconv\ signal is shown in Figure  \ref{fig:ana_step2_sindrum_positron_best_fit_signal}
together with the SINDRUM-II positron data and the RMC background. 
The RMC normalization comes from the fit in the region p < 88 MeV/c, explained in Section
\ref{sec:kmax_determination}. To guide the eye, the $\mu^- \ra e^+$ conversion signal is
normalized to 20 events.

The excess of events above 88 MeV/c has a shape consistent with the shape
of the expected $\mu^- \rightarrow e^+$ signal; however, the average momentum of the
data events in the group is about 1 MeV/c lower than expected from the signal.
The estimated statistical uncertainty on the position of the center of gravity of those
events is 0.24 MeV/c, so the expected position of the $\mu^- \rightarrow e^+$ conversion
signal is about 4$\sigma$ higher than calculated from the data. An attempt to fit the data
in the range [89, 92] MeV/c  with the function having the shape of the  $\mu^- \rightarrow e^+$
conversion signal returns a p-value of 0.004.
This strongly disfavors the exotic ($\mu e$ conversion) interpretation of the excess.

The track charge mis-identification probability in the SINDRUM-II detector
is about 0.2\% \cite{sindrum_ii:Kaulard1998}, which can't explain the excess
of positron events.
%

\section{An exclusive  RMC transition?}

As it has already been mentioned, the excess of the positron events in the tail
could be explained by the contribution of RMC accompanied by a nuclear transition
from the ground state of \Au{197} to an exclusive final state of $^{197}\rm Pt$.
As the ground state of \Au{197} has spin parity $3/2^+$ and the ground state of $^{197} \rm Pt$
has spin-parity of $1/2^-$, an allowed dipole transition between the two states
could result in mono-energetic photons with the energy E = 94.3 MeV.
A uniform, in a first order approximation, distribution of the electron-positron energy
splitting could result in a flat contribution extending up to p = 93.2 MeV/c
added to the rapidly falling measured positron spectrum from converted RMC photons.

In total, the SINDRUM-II positron spectrum on a gold target has 456 events with p > 77 MeV/c.
14 out of 456 events have p > 88 MeV/c. Attributing one event - with the highest
momentum - to cosmics/RPC, there are 13 events left with reconstructed positron
momenta in the region 88-93 MeV/c.
Assuming that all of them are due to an Au(GS) $\ra$ Pt(GS) RMC transition and a flat positron
spectrum from 0 - 93 MeV/c, we estimate the total number of such events
as $13\text{ events}/5\text{ MeV/c}\cdot 93\text{ MeV/c} \approx 250$ events.

\begin{figure}
  \hspace*{-1.3cm}%
  \begin{tikzpicture}
    \node(base) at (5,0){};
    \node[anchor=south west,inner sep=0] at (0.0,0.) {
      \includegraphics[width=0.6\textwidth]{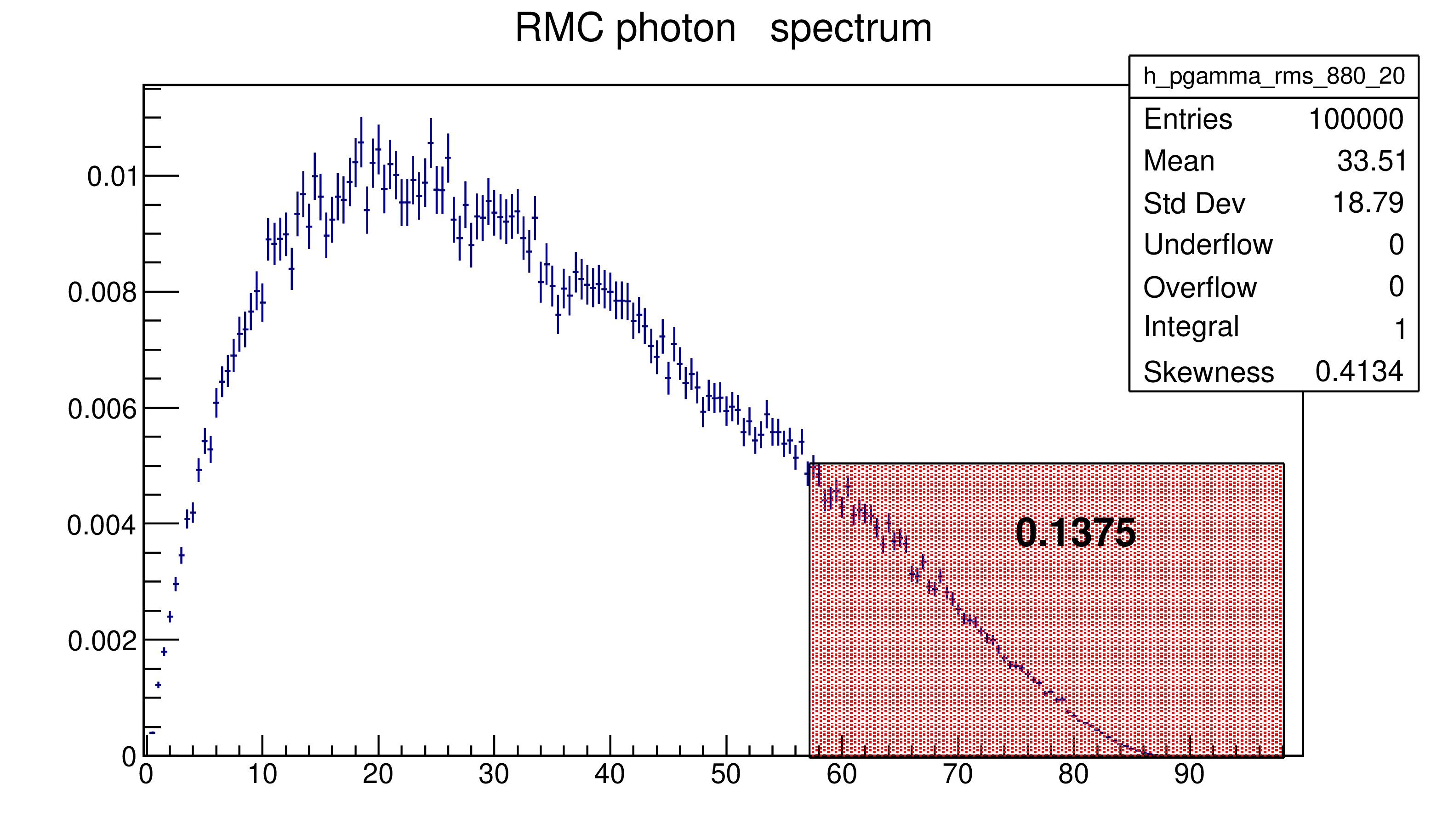}
    };
    \node[anchor=south west,inner sep=0] at (8.1,0.0) {
      \includegraphics[width=0.6\textwidth]{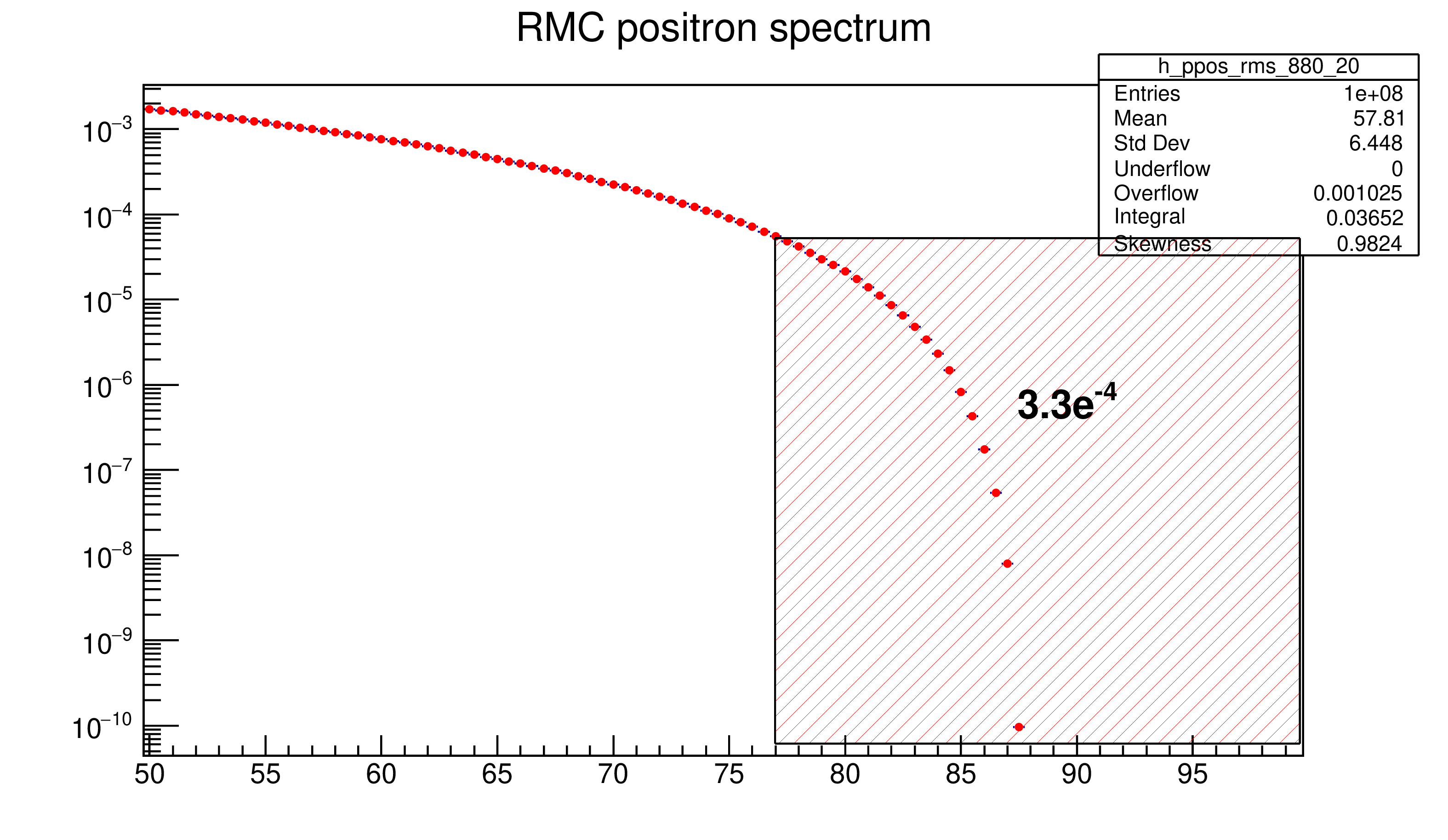}
    };
  \end{tikzpicture}
  \caption {
    \label{fig:rmc_photon_and_positron_spectra}
    RMC photon (left) and $e^+$ (right) momentum spectra, $\kmax = 88$ MeV.
  }
\end{figure}

For an RMC spectrum with $\kmax = 88$ MeV, shown in Figure ~\ref{fig:rmc_photon_and_positron_spectra}, 
about 14\% of photons have energies above 57 MeV.
The TRIUMF RMC spectrometer measured 2000-3000 data events per target \cite{Bergbusch:1999ms}.
For an experimental energy cutoff E > 57 MeV, that translates into 15,000 - 20,000
events in the whole spectrum. 

Similarly, for $\kmax= 88$ MeV, about $3.3\cdot 10^{-4}$ of all RMC positrons have
p > 77 MeV/c. Assuming a momentum-independent efficiency, 442 reconstructed events
with p > 77 MeV/c corresponds to $1.3\cdot 10^6$ RMC events in the full spectrum
and $1.9\cdot 10^5$ RMC events with photon energies above 57 MeV. This is two orders
of magnitude higher than the per-target statistics of the TRIUMF RMC spectrometer.

The branching ratio of the exclusive transition could be estimated at 
$\sim 2.5\cdot 10^2/1.3\cdot 10^6 ~ \sim ~ 2\cdot 10^{-4}$, so for typical
per-target statistics of \cite{Bergbusch:1999ms}, one would expect such a transition
to contribute $2\cdot 10^4\cdot 2\cdot 10^{-4} ~\sim 5$ events.
Given the photon energy resolution of FWHM $\sim$ 7 MeV, it would be rather difficult
for the TRIUMF RMC spectrometer to resolve transitions of that strength in the measured
spectra. However, a measurement with a photon energy resolution of about 1 MeV and
statistics an order of magnitude higher than that of \cite{Bergbusch:1999ms}
would allow one to see such a transition.

\newpage
\section{ Summary }
In their 2006 paper which sets the current best limit on \mumemconv\ on a gold target
\cite{sindrum_ii:Bertl2006}, the SINDRUM-II collaboration published, along with
the electron momentum distribution, the momentum distribution of reconstructed positrons.
Near the positron spectrum endpoint, there is an excess of events which has not
been discussed by the authors.

To understand the origin and potential implications of this excess, we developed
a simple detector response model, tuned it using the SINDRUM-II electron data,
and used the tuned model to describe the positron data. The excess of high-momentum
events in the positron data is statistically significant, comprised of about 13 events
with the background expectation of about 1.4. The background expectation is dominantly
due to RPC and cosmics, the RMC background above 88 MeV/c is < 1 event. 

Interestingly, the excess has a width consistent with the SINDRUM-II detector resolution.
The expected position of the $\mumepconv[Au][Ir]$ signal is 1 MeV/c, or $\sim4\sigma_p$,
higher, and fitting those events with the $\mu^- \ra e^+$ signal shape has a p-value of 0.004;
this strongly discourages the exotic interpretation.

The excess could be due to an exclusive dipole RMC transition between \\ 
$^{197} \rm Au(GS)$$\ra$$^{197} \rm Pt(GS)$, with a branching fraction of about
$2\cdot 10^{-4}$. Given the statistics and resolution of the published RMC 
measurements, such a transition would not be resolved experimentally.

The observation has significant implications for the upcoming searches 
for processes of $\mumepconv$ and $\mumemconv$ conversion. In the $\mumepconv$ channel,
exclusive RMC transitions could simply fake the signal; in the $\mumemconv$ channel,
those transitions could significantly distort the predicted SM background shape.

To fully exploit the physics potential of experiments such as Mu2e and COMET,
a better theoretical understanding of the endpoint of the RMC spectrum on nuclei
is needed. A high-resolution measurement of the RMC photon spectra needs to be carried
out and compared to the theoretical predictions. Without that, the sensitivity
of the upcoming searches might be limited by the unknown probabilities of RMC
transitions to the exclusive low lying states of the daughter nuclei.

\section{ Acknowledgments }

We want to thank our colleagues at Mu2e, who offered advice for this study.
We also appreciate the Velasco group at Northwestern for encouraging M. MacKenzie to
participate in these studies. His research was in part funded by the ``Research in the Energy,
Cosmic and Intensity Frontiers at Northwestern University'' award,  DE-SC0015910.
Work of P.Murat was supported by the Fermi National Accelerator Laboratory, managed and operated by Fermi Research Alliance, LLC under Contract No. DE-AC02-07CH11359 with the U.S. Department of Energy. The U.S. Government retains and the publisher, by accepting the article for publication, acknowledges that the U.S. Government retains a non-exclusive, paid-up, irrevocable, world-wide license to publish or reproduce the published form of this manuscript, or allow others to do so, for U.S. Government purposes.

\appendix
\section{Calculation of the positron energy in $\mumepconv[Au][Ir]$ conversion}
We consider only the ground state conversion process. 
For this, the conversion energy is given by:
$$
E_{e^+} = m_\mu - B_\mu - E_{recoil} - \Delta_{Z-2}
$$

\noindent where $m_\mu$ is the muon mass, $B_\mu$ is the binding energy of the 1s energy level,
$E_{recoil}$ is the kinetic energy of the recoiling nucleus, and $\Delta_{Z-2}$ is the difference
between the incoming and outgoing nuclear masses. 

\begin{table}[h]
  \begin{center}
    \begin{tabular}{|l || c |}
      \hline
      Parameter & Value  \\
      \hhline{|=||=|}
      $m_\mu$ & 105.6583745(24) MeV/c$^2$ \cite{PDG}\\
      \hline
      $m_e$   & 0.5109989461(31) MeV/c$^2$ \cite{PDG}\\ 
      \hline
      $1u$ & 931.49410242(28) MeV/c$^2$ \cite{PDG}\\ 
      \hline
      $B_\mu$ & 10081.23 keV \cite{MuonBindingEnergies:1974}\\ 
      \hline
      $A_r(\Au{197})$ & 196.96656879(71) u  \cite{NuclearData:2017}\\
      \hline
      $M_N(\Au{197})$ & 183432.828(1) MeV/c$^2$\\
      \hline
      $A_r(\Ir{197})$ & 196.969 655(22) u  \cite{NuclearData:2017}\\
      \hline
      $M_N(\Ir{197})$ & 183436.725(21) MeV/c$^2$ \\
      \hline
      $\Delta_{Z-2}$ & 3.897(21) MeV/c$^2$ \\
      \hline
    \end{tabular}
  \end{center}
  \caption{Parameters used in the $\mumepconv[Au][Ir]$ positron energy calculation.}
  \label{table:parameters}
\end{table}

Since the muon orbit is $\sim$ 200 times closer to the nucleus than the electrons',
one can assume the electrons do not contribute to the muon capture process. Therefore,
the electron masses are not considered and instead only the nuclear masses are used:
$M_N = M_A-Z\cdot m_e$, where $M_A=A_r\cdot u$, $A_r$ is the relative mass, 
u is the atomic mass unit, and $m_e$ is the electron mass.

The recoiling energy is given by considering the two body decay: $N(\mu^-+\Au{197}) \ra e^+ + \Ir{197}$.
Considering the rest frame of the muonic gold atom, the decay products must satisfy
$p_{e^+} + p_{^{197}Ir} = p_{o} = 0$. The energy of the iridium nucleus is then given by:
$$
E_{^{197}Ir} = \frac{M^2 + M_N(\Ir{197})^2 - m_e^2}{2M} = E_{recoil} + M_N(\Ir{197})
$$
where M is the mass of the muonic gold nucleus system, $M=M_N(\Au{197}) + m_\mu - B_\mu = 183528.405$ MeV/c$^2$.
Using the parameters in Table \ref{table:parameters}, we get $E_{recoil} = 0.023$ MeV.

Combining these together, the $\mumepconv[Au][Ir]$ positron energy for the ground state
transition is $E_{e^+} = 91.657(21)$ MeV, where the uncertainty on $B_\mu$ is assumed to be $\sim$1 keV.



\addcontentsline{toc}{section}{Bibliography}
\begin{thebibliography}{99}                 
\bibitem{sindrum_ii:Kaulard1998}
  Kaulard, J. and others (SINDRUM II), Phys. Lett. B, {\bf 422}, 334 (1998)

\bibitem{sindrum_ii:Bertl2006}
  W. H. Bertl et al. (SINDRUM II), Eur. Phys. J. C {\bf 47}, 337 (2006).
  

\bibitem{Watanabe:1993}
  R. Watanabe, K. Muto, T. Oda, T. Niwa, H. Ohtsubo, R. Morita, and M. Morita,
  Atomic Data and Nuclear Data Tables, {\bf 54}, 1, 165 (1993)

\bibitem{RoodTolhoek:1965} 
  H.P.C. Rood and H.A. Tolhoek, 
  Nuclear Physics, {\bf 70}, 3, 658 (1965) 

\bibitem{sindrum_ii:Kaulard1997_Thesis}
  Kaulard, Joerg, PhD Thesis, RWTH, Aachen, unpublished, in German (1997)

\bibitem{NuclearData:2017}
  G. Audi, F.G. Kondev, Meng Wang, W.J. Huangand,  and S. Naimi,
  The NUBASE 2016 evaluation of nuclear properties, 
  Chinese Physics C, v41, 030001 (2017)

\bibitem{Bergbusch:1999ms}
  Bergbusch, P.C. et al, Phys. Rev. C, {\bf 59}, 2853 (1999)

\bibitem{PDG}
  P.A. Zyla et al. (Particle Data Group), Prog. Theor. Exp. Phys. {\bf 2020}, 083C01 (2020).
  
\bibitem{MuonBindingEnergies:1974}
  R.Engfer, H.Schneuwly, J.L.Vuilleumier, H.K.Walter and A.Zehnder,
  ATOMIC DATA AND NUCLEAR DATA TABLES {\bf 14}, 509 (1974)
\end{thebibliography}
\end{document}